# Unprecedented three-dimensional magnetic cloak working from DC to 250 kHz

Jianfei Zhu[1,*], Wei Jiang[1,*], Yichao Liu[1], Ge Yin[1], Jun Yuan[1], Sailing He[1,2,‡] and Yungui Ma[1,‡]

[1]State Key Lab of Modern Optical Instrumentation, Centre for Optical and Electromagnetic Research, College of Optical Science and Engineering, Zhejiang University, Hangzhou 310058, China

[2]Department of Electromagnetic Engineering, School of Electrical Engineering, Royal Institute of Technology, S-100 44 Stockholm, Sweden

*These authors contributed equally to this work.
‡E-mail: yungui@zju.edu.cn or sailing@kth.se

**Invisible cloaking is one of the major advancements of modern electromagnetism assisted by the development of metamaterials and transformation optics (TO). However, the practical potential of the generally transformed device, in particular for high frequencies (e.g., microwave to visible light), is fatally challenged by the complex material parameters they demand. On the other hand, it will be advantageous and also technologically instrumental to design cloaking devices for applications at low frequencies where electromagnetic (EM) components are favorably uncoupled. In this work, we vastly develop the bilayer approach to create a three-dimensional (3D) magnetic cloak able to work in both static and dynamic fields. Under the quasi-static approximation, we demonstrate a perfect magnetic cloaking device with a large frequency band from zero to 250 kHz. The practical potential of our device is experimentally examined by using a commercial metal detector. Considering its simplicity and efficiency, our bilayer architecture may encourage research efforts to find a first real cloaking application that requires shielding of magnetic field without disturbing it.**

In the last decade, invisible cloaking or hiding things from detection has gained extensive research enthusiasm in the field of metamaterials and pertinent communities since the proposal of the original conceptions[1,2]. Modern TO techniques have been developed to configure and modulate the cloaking devices associated with the development of complex metamaterials with extreme EM properties[3-8]. In order to improve the practical feasibility, various modified versions for EM cloaks have been proposed and demonstrated, among which the most explored are the quasi-conformal[9-16] and (later) the bilinearly transformed cloaks[17-19]. These modified TO approaches allow the usage of weakly anisotropic or isotropic artificial/natural materials to fabricate the appliances, which greatly improves the loss and bandwidth properties. However, individually they have their own inherit limits either in incident angles or in polarization states[20,21]. Precise fabrication of these devices with acceptable reproducibility and cost is still very challenging, especially for those aimed to work at high frequencies (i.e., microwave to terahertz or visible light). These hurdles need to be overcome in the future with smarter designs or more accessible cloaking algorithms.

On the other hand, invisible cloaks for single components of EM fields at direct current (DC) or low frequencies are also of practical importance[22-29], especially for magnetic fields that are universally involved in numerous modern facilities and technologies[22-24,27-29]. Superconducting (SC) materials with static perfect diamagnetism ($\mu = 0$) play a key role in designing a DC magnetic cloak[22,30,31]. Technically though, it won't be easy to fabricate a transformed magnetic cloak even by a composite of SC and ferromagnetic (FM) materials. To solve this issue, a promising bilayer approach was initially proposed by Sanchez's group to build a static magnetic cloak on the basis of uncoupled Maxwell's equations, i.e., Laplace's equation[23]. Although the theory of this bilayer approach starts from a uniform external field, it is found to still be functional in certain inhomogeneous environments, even like a magnetic dipolar field[28]. Because of its simplicity and efficiency, the bilayer cloak approach has attracted considerable attention and has been quickly applied to many other parallel systems mainly controlled by diffusion equations, such as thermal flux[32,33], electric current[34], ions[35] and diffusion light[36]. One application in particular is an electro-thermal bifunctional cloaking device, which was recently demonstrated by the authors on the basis of this bilayer architecture[37].

These experimental progresses enabled by the bilayer approach may bring us more confidence to pursue a practical cloaking device specifically for applications at low frequencies where the quasi-static approximation is valid[38]. Using a torus made of SC and FM composites, Sanchez's group has demonstrated the possibility to realize a static magnetic cloak firstly in a two-dimensional (2D) field[23]. A real practical device is generally required to work in a 3D space. More importantly, dynamical functionalities with selectable operation bands are highly desired, especially corresponding to those used by magnetic field induction (MFI) technologies that are commonly relied on to uncover the hidden or underground magnetic/metallic objects[39]. In another paper[27], Sanchez's group made a first attempt in this aspect and showed a limited quasi-static magnetic cloaking effect from zero to 144 Hz utilizing a similar 2D cloaking structure. The sample fabrication will become significantly more difficult in three dimensions because the usual flat FM or SC sheets cannot be applied here anymore for spherical topologies. Dynamic field applications will also raise the requirements on material properties such as non-linear hysteresis and/or conductive loss of the FM component and the transport capability or critical field strength of the SC component. Besides these material issues, there is a general question of whether the SC component can still be as effective as a zero-permeability entity in time-varying fields where inhomogeneous vectorial Helmholtz equations have to apply.

Regarding these critical issues, in this work, we vastly develop the bilayer approach to pursue a magnetic cloak operational in a 3D quasi-static field by optimizing material properties. In contrast with the metallic alloys used before[23,27,28], resistive high-quality ferrite is employed here to remove the eddy-current loss and more importantly to acquire a linear magnetic response in a relatively broad field range. For the magnetized sample, a nearly flat permeability spectrum is achieved in a frequency band from DC to hundreds of kilohertz. The SC component we use is carefully manufactured from single crystal Yttrium barium copper oxide (YBCO) cylinders, whose bulk and single crystal features could exclude many possible negative material issues related to inductive loss. With such a bilayer structure, experimentally we demonstrate a perfect 3D magnetic cloaking effect from DC to a maximum measurement frequency of 250 kHz, which covers the operation bands of nearly all EMI appliances. The application potential to hide objects in a real field is experimentally examined by using a commercial metal detector. Considering the structural simplicity and practical efficiency, this bilayer magnetic cloak is a strong candidate for a first real application in the field where the influence to external magnetic fields needs to be minimized or totally eliminated.

**Results**

**Sample design and fabrication.** Figure 1 provides a schematic of the bilayer structure consisting of the SC inner shell (black) and FM outer shell (brown) in a dielectric background. Each shell consists of two firmly connected identical halves. The SC shell (inner radius $R_1$ and outer radius $R_2$) was machined and etched from two YBCO single crystal cylinders. In Cartesian coordinates, the $z$-axis is defined along the $c$-axis of the YBCO's unit cell and the $xy$ plane is parallel with the $ab$ lattice plane. In this manner, the maximum applicable magnetic field is different along the $z$-axis than a direction in the $xy$ plane due to the material anisotropy[40]. The FM shell of outer radius $R_3$ is a composite of NiZn soft ferrite powders and paraffin matrix by a proper weight ratio. The fabrication details can be found in the Method section. Assuming a uniform static external field and a perfect SC shell (skin depth or London penetration depth is in the sub-micron scale[41]), the FM component required by an ideal 3D magnetic cloak should have a permeability of (see Ref. [23] and also the supplementary note 1)

$$\mu_{FM} = \frac{2R_3^3 + R_2^3}{2R_3^3 - 2R_2^3}. \quad (1)$$

By this bilayer structure, any magnetic or conductive object (yellow) placed inside the cloaked region (white) should be magnetically invisible (or undetectable) to a nearby observer [29]. In practice, we select $R_3 = 1.5R_2 = 15$ mm so that the desired $\mu_{FM}$ is equal to a small number (1.63). The relatively large thickness and low permeability of the FM shell is helpful in minimizing the demagnetization shape effect and thus improves the isotropic response of the FM shell. The diluted magnetic ingredient in the composite also helps control the magnetic loss.

**Simulation results**. Numerical simulation is first carried out to examine the cloaking performance of the proposed device by COMSOL multiphysics. The modeling details for static and dynamic simulations can be found in the Method section. Figures 2a-2c show the static magnetic field patterns at a cross-section passing the center of the samples made of SC only, FM only and the bilayer composite (SC+FC), respectively. The normalized strength of the magnetic field is represented by different colors, and the field direction is represented by the black arrow lines. It is apparent that the magnetic force lines on the top of the sample are expelled by the sole SC shell (Fig. 2a) and concentrated by the sole FM shell (Fig. 2b), while these perturbations are completely canceled by their proper combination (Fig. 2c). These results are more quantitatively illustrated by the field change curves calculated along one straight line at a 5mm distance above the sample in the supplementary Fig. S1a.

To simulate the dynamic response, we assume our single-crystal YBCO bulk has a conductivity $\sigma = -i/(\omega\mu_0\lambda_L^2)$ S/m at 77 K determined by the normal-state eddy current[41,42], where $\omega$ is the angular frequency, $\mu_0$ is permeability and $\lambda_L$ is the London penetration depth. As a perfect conductor, SC's dynamic behavior is dominantly decided by its electric properties, and the influence of permeability is less important and practically could be neglected. Figures 2d-2f show the simulated snapshots of the real magnetic field distributions at 25 kHz for the devices consisting of different components. They exhibit similar field interaction patterns as their static counterparts shown in the top panel. These dynamic behaviors are nearly independent of the simulation frequency from 100 Hz to 1 MHz. Under the quasi-static approximation, these frequency behaviors are quite reasonable because Maxwell's equation outside the SC region could still be effectively approximated by a Laplace equation, and the infinitely large conductivity of a superconductor will take over the static Meissner effect to screen out the oscillating EM field. The latter effect makes the SC component still act as a perfect diamagnetic entity. Consequently, at low frequencies, an SC or FM shell alone or their combination will almost replicate their static field behaviors, provided that their material parameters are constant at different frequencies. The perfect dynamic cloaking performance is evidenced by the magnetic field distribution in Fig. 2f. A more quantitative analysis is given in the supplementary Fig. S1b by plotting the relative change of the nearby field.

**Cloaking measurement in static magnetic fields.** For the static field measurement, we use a pair of commercial Helmholtz coils that can provide a maximum magnetic field 25 mT. All the samples are first cooled down in a liquid nitrogen bath (77 K) with no external magnetic field before measurement. To check the field perturbation, we scan the local field distribution along the line $z = R_3 + h$ in the $xz$ plane, where $h$ (= 5 mm) is the proper distance to the north pole of the FM shell. The relative change percentage, defined by $100(V_s-V_0)/V_0$ with $V_s$ and $V_0$ representing the measured signals with and without the sample, respectively, is used in our descriptions to evaluate the field variation due to the introduction of an object. The relative change for ideal cloaking should be zero. Figure 3a plots the relative change curves of the magnetic field measured for the bilayer sample (squared black) and the two references with an SC (dotted red) or FM (triangle blue) shell only. In the measurement, a uniform external field of 2.5 mT is applied along the $z$-axis. The concentrating or expelling disturbance to the nearby magnetic field is clearly observed due to the existence of the FM or SC shell, respectively. Note as a near-field effect this influence manifests only within a deviation distance of 35 mm, which agrees well with the simulation given in Fig. S1a. In contrast, the relative field change caused by the bilayer sample is substantially suppressed and the measured amplitude fluctuates around null (dashed line), with absolute values within an uncertainty error of 0.3% of our experiment system. Because the sample has a rotational symmetry, similar field patterns could be expected in other positions. Therefore, it can be concluded that our bilayer structure gives a reasonably good realization of a DC magnetic field cloaking effect in 3D space.

For practical applications with different purposes, the field strength tolerance is a very important factor in evaluating the capability and potential of a real cloak. In this aspect, the linear response capability of the FM component and the maximum critical field for the SC component need to be carefully examined. Regarding the first issue, NiZn spinel ferrites with relatively large magnetocrystalline anisotropic energy among soft magnetic materials have been employed in this work in order to acquire a relatively high-field linear magnetization property. The supplementary Fig. S4a plots the magnetization hysteresis loop of our FM composite at a maximum field of 1 T at 77 K. The inset for a zoom-in minor loop shows the magnetic composite has a good linear magnetization behavior at least up to 20 mT, which could be large enough for most low-field applications. This linear characteristic is important for the current amplitude and frequency based field perturbation measurement[29].

Figure 3b gives the relative changes of the magnetic field as a function of the external field strength measured at a fixed point $z_0 = R_3 + h$ on the $z$-axis for the structures with different components. For the sole FM shell, the relative field change (blue) is nearly constant at ~6%, which is consistent with the linear magnetization loop given in the inset of Fig. S4a. For the samples with a SC shell, we repeat the measurement by applying an external magnetic field along the $x$- and $z$-axis, with $H$ parallel to the $c$-axis of the YBCO unit cell and with $H$ perpendicular to it, respectively. For the SC and bilayer samples, the relative field change (red) measured along these two different directions starts to "split" at a field near 2.8 mT. As a typical type-II superconductor, YBCO's lower critical field in the $ab$-plane (~ 2.5 mT) has been found to be nearly ten times smaller than that along the $c$-axis (~ 26 mT) at 77 K[39]. Below this lower limit, it works in the Meissner state as a perfect diamagnet and turns into a mixed state above this limit where the field starts to penetrate the SC body. It imposes our magnetic cloak an upper limit for the working field (no more than 2.5 mT) for general applications. However, this limit could be improved if the external field direction is known in advance.

**Cloaking measurement in dynamic magnetic fields**. For dynamic measurement, we use a pair of custom-made Helmholtz coils with proper inductance. A Stanford signal generator is used to excite the coils, and the magnetic field near the sample is inductively measured through a small copper loop connected to a lock-in amplifier.

The field exposed to the sample is less than 1 mT, and the measurement frequency varies from 5 Hz to 250 kHz to suit the band of our lock-in amplifier. Figure 4 gives the measured relative change of the harmonic magnetic field intensity at a fixed point $z_0 = R_3 + h$ on the $z$-axis for various samples. From the curves, we see that the dynamic responses of these samples are nearly independent of the operating frequency. The inset of Fig. 4 shows the measured room-temperature permeability spectrum of our FM composite from 100 Hz to 250 kHz. A nearly constant permeability around 1.54 is observed, which is essential for the large bandwidth of our cloaking device. Permeability at the working temperature (77 K) is estimated by multiplying the ratio of the saturation magnetization measured at these two different temperatures. Fine-tuning of the composition in experiment is carried out till the lowest field perturbation is achieved. The measured relative changes for the SC and FM shells only remain at about 6% and 7%, respectively, while that for the bilayer sample is desirably diminished to nearly zero (absolute amplitude < 0.02%). Note the measurement at low frequencies (< 50 Hz) has an increased uncertainty error due to the fact that the measured induction signal is proportional to the operation frequency. However it will not impact the conclusion that the measured results verify the dynamic cloaking performance of our device from low frequency up to 250 kHz, where the quasi-static approximation works fairly well.

To get a full picture of the perturbation to the exciting field, we measure the relative change of the field along the straight-line at $z = R_3 + h$ in the $xz$ plane at 25 kHz by two ways: one is scanning the probe over a fixed sample and the other is moving the sample over a fixed probe. These two measurements examine the capability of our bilayer sample to work in a uniform field and a space-varying field, respectively, and the results are given in Figs. 5a and 5b, respectively. The latter condition caused by the real field distribution of our finite exciting coils corresponds to a maximum relative incident field change of about 4% along the measured line. The disturbance to the external field by the SC or FM shell due to magnetic induction or moment is clearly observed. For the bilayer sample, the disturbance to the near field is substantially suppressed, and the measured relative change is smaller than 0.2%. These measured results are in good agreement with the simulations shown in Fig. S1b.

So far we have confirmed the dynamic cloaking capability of our bilayer structure in time-harmonic uniform magnetic fields with frequencies varying from zero to 250 kHz. Through a Fourier transform we may expect that the same device can also work in the time domain as well with an arbitrary time-varying field, e.g., an impulse field. In addition, it was numerically predicted that the performance of the bilayer cloak is robust against the exciting field distribution[28], which is similarly verified in our electric-thermal bifunctional cloak also designed by the bilayer approach[37]. In the last experiment here, we check the time-space response of our sample by using a commercial metal probe scanner that detects a conductive object by perceiving the change of the EMI signal[43]. The scanner has a transmitter coil to generate the probing field at a working frequency of 25 kHz and a pick-coil to measure the field change. In experiment, it is held to sweep over the sample at a distance of about 2 cm and will respond to the EM signature of a conductive object in two manners, i.e., flashing the indicator (green to red) and making a warning sound 'beep'. The measured results are given in the supplementary Movies S1 and S2. Movie S1 shows that the scanner can perceive the room-temperature bilayer sample (middle) and also two nearby references, i.e., a metallic badge (left) of Zhejiang University and an aluminum sphere ($R$ = 10 mm) coated by 5-mm-thick paraffin (right), as shown in the supplementary Fig. S5. After immersed in liquid nitrogen for several minutes, as shown in Movie S2, our cooled bilayer sample successfully 'cheats' the prober without triggering the alarms. In this case, our bilayer shell can hide a metallic object inside and make the whole structure transparent to dynamic magnetic fields. The exact voltage signals generated by the pick-up coil are also measured by an oscilloscope with the results plotted in the supplementary Fig. S6. From the modulation of the signal amplitude, we could recognize an approaching uncovered metallic object but are unable to access the cloaked sample.

**Discussions**

We have experimentally confirmed the functionality of the bilayer cloak for both DC and dynamic fields under the quasi-static approximation. This assumption is valid only if the implemented device has a perfect SC component with a negligible field penetration depth compared to the sample size. An inner shell made of a common metal that has a frequency-dependent penetration depth at our desired frequencies cannot mimic the behavior of a perfect diamagnet to completely screen out the primary field[44]. The supplementary Fig. S2 shows that the cloaking effect disappears when the inner SC shell is replaced by a conductor like copper. For our bilayer structure, the FM and SC components could be regarded as two polarization-opposite dipoles that balance each other's influence and induce a magnetic transparency effect[45]. In principle, several microns could be thick enough for the single-crystal SC shell if its manufacture was feasible.

To obtain good magnetic cloaking performance, the precise fabrication of the FM component with the desired properties is another necessary condition. First, the permeability of the FM component has to be accurately realized. The supplementary Fig. S3 shows that small deviations in the permeability from the optimal value by 6% will change much of the overall cloaking effect. Experimentally we reach the optimal value by gradually tuning the composition of the FM component around an estimated composition. The linearity of permeability within 20 mT is achieved by employing relatively 'hard' NiZn spinel ferrite. Another issue for dynamic cloaking is the magnetic loss, which origins from both hysterical and residual losses for our insulating composite. In experiment, the magnetic field amplitude we apply is less than 1 mT and under this condition, the magnetization switching around the remnant state is mostly a reversible process, and the residual loss due to the phase lag of magnetization rotation will be the dominant loss[46]. However, it won't be a serious problem in our bilayer structure because the ferrite constituent is vastly diluted by the paraffin matrix, which leads to a very small averaged imaginary permeability (< 0.02), as shown in supplementary Fig. S4b. For strong external fields, the weight of the hysterical loss will increase as theoretically predicted by Rayleigh's law[23]. On the other hand, this magnetic nonlinear loss could be potentially controlled by adopting a lower loss material composed of isolated single-domain particles (i.e. removed Barkhausen jump) such as $Co_2Z$ hexagonal ferrite[47], whose higher anisotropy also helps enlarge the maximum field limit and the operation bandwidth as well. By the current architecture, the maximum working frequency of up to 1 MHz can be expected. Beyond this, the magnetic cloak effect will gradually disappear due to the permeability change of the FM component and also the failure of the quasi-static approximation.

**In conclusion**, we have demonstrated a 3D quasi-static magnetic cloaking device able to work from DC to 250 kHz. Elegant control of the electric and magnetic material losses is realized by selecting high-quality NiZn ferrite and single-crystal YBCO as the ingredients of the FM and SC components, respectively. An isotropic flat permeability with negligible residual loss is obtained from DC to hundreds of kilohertz for our FM composite. The perfect SC shell helps exclude many possible negative material issues related to inductive loss. In the experiment, the dynamic performance of our bilayered 3D cloak in a wide frequency band from DC to 250 kHz has been verified from the nearby field perturbation measurement, which is in good agreement with the simulations. The practical potential of our device in anti-probe technology for magnetic fields is also experimentally examined using a real metal detector. The same architecture may also find essential applications in places such as bio-experiments or magnetically sensitive equipment where perfect screening without distorting external magnetic fields is specifically required.

**Acknowledgements**
The authors are grateful to the partial supports from NSFCs 61271085, 91130004, and 91233208, NSFC of Zhejiang province LR15F050001, the Program of Zhejiang Leading Team of Science and Technology Innovation (2010R50007) and the support by the Fundamental Research Funds for the Central Universities.

**Author contributions**
The samples were fabricated by J.F.Z. and measured by J.F.Z. and W.J. Modeling is conducted by Y.G.M. and Y.C.L. VSM and permeability analysis is performed by G.Y. and J.Y. The work is under supervision by Y.G.M. and S.L.H. All the authors contributed to the writing and discussions of the manuscript.

**Additional information**
Supplementary information is available in the online version of the paper. Reprints and permissions information is available online at www.nature.com/reprints. Correspondence and requests for materials should be addressed to Y.G.M. and S.L.H.

**Competing financial interests**
The authors declare no competing financial interests.

**Methods**

**Simulation of the sample.** Both static and quasi-static simulations are performed using the axial symmetry module of COMSOL multiphysics. The 3D sample is constructed with structural parameters $R_1 = R_2/2 = R_3/3 = 5$ mm and materials parameters $\mu_{FM} = 1.63$ for the FM component and $\sigma = -i/(\omega\mu\lambda_L^2)$ S/m for the single-crystal SC component[41,42]. We take a relative large penetration depth $\lambda_L = 1$ μm. A Laplace equation for the scalar magnetic potential is built to simulate the static field behaviors of the samples defined in a rectangular domain formed by two pairs of constant potential and magnetic insulation boundaries. For the dynamic simulation, an RF model with scattering boundary conditions has been based to mimic the quasi-static situation. A tall cylindrical current surface with height and radius far larger than the FM shell's radius $R_3$ has been introduced to yield a well-defined uniform excitation environment for the sample. Since we only care about the near-field effect, the inaccuracies appearing in the far-field locations for low-frequency simulations could be practically excluded from our consideration.

**Fabrication of the sample.** YBCO superconductor is used in our work due to its resilience at high temperatures. Two identical halves of the SC component with inner radius $R_1$ and outer radius $R_2$ are machined and chemically etched from two commercially bought YBCO cylinders (diameter = 20 mm and height = 10 mm). Great care and patience are required during the fabrication process due to the very fragile nature of YBCO single crystals, which occupied most of the sample preparation time. The FM component with inner radius $R_2$ and outer radius $R_3$ is made of commercial MnZn ferrite powders and paraffin by using a standard molding process. Magnetic hysteresis loops of the composite are measured at both room temperature and 77 K by a vibrating sample magnetometer (VSM Lakeshore 7400). Reducing the temperature increases the measured saturation magnetization ($Ms$) by about 6%. The effective relative permeability of the composite as a function of the weight ratio of these two ingredients is first measured at room temperature by using an impedance analyzer (Agilent 4294A) for donut-shaped samples. We estimate the permeability at 77 K by multiplying a factor from the change in $Ms$. Finer tuning of the weight ratio around a desired estimated value is conducted till we experimentally achieve an optimal cloaking effect. For the results reported here, the weight ratio of the powders to paraffin is 0.85, which corresponds to a room-temperature relative permeability at about 1.54±0.01 from 100 Hz to 0.5 MHz. It is important to note that our magnetic sample for both static and dynamic experiments works in the linear portion of the hysteresis loop, and the non-zero remnant magnetization provides a self-biasing field.

**Measurement.** For the static field measurement, the samples were placed at the center of a pair of commercial Helmholtz coils (diameter = 36 cm and spacing = 14.5 cm) which can provide a maximum static magnetic field of 25 mT. Before measurement, the samples were first cooled down with liquid nitrogen to 77 K with no external field. A low-temperature Hall sensor (Lakeshore HGCT-3020) was used to scan the z-component of the magnetic field along one straight line at $z_0 = R_3 + 5$mm in the $xz$ plane, i.e., above the sample by a 5mm distance. High reproducibility of measurement data is guaranteed by utilizing a programed step-motor to move the prober. For the dynamic field measurement, a pair of Helmholtz coils (diameter = 33 cm, spacing = 11 cm and turns $N = 30$) was built, which can satisfy the frequency band measurement, driven by a signal generator (Standard DS345). The harmonic field intensity near the sample is inductively measured by a ring prober (diameter = 4 mm) made of 80-turn copper wire loops, connected to a lock-in amplifier (Signal Recovery 7270). The dynamic frequency is measured from 5 Hz to 250 kHz, decided by the lock-in amplifier we use. The voltage $V$ used in this work is a root-mean-square value of the measured oscillating signal. The induced signal is proportional to the frequency and suffers an increased measurement error at low frequency. For the curves given in Fig. 5b, the prober is fixed at $z_0 = R_3 + 5$mm and the samples are displaced along the $x$-axis by ±50mm intervals. In this manner, the samples experience a relative change of the incident field strength of about 4%. To examine the practical potential, a commercial handhold metal detector (Tianxu TX1001B) is used to scan the samples over a distance of roughly 2 cm. It consists of a pair of transmitting and receiving coils as well as an electronic circuit, and works at 25 kHz. When swept over a conductive or magnetic object, the EMI signal of the receiving coil will change in intensity and phase, which will then trigger the alarm (flashing the indicator and making a sound) if the change is larger than the threshold. The exact voltage signal of the receiving coil with respect to time is also measured by an Agilent (34410A) oscilloscope while keeping the scanner sweeping back and forth over the samples.

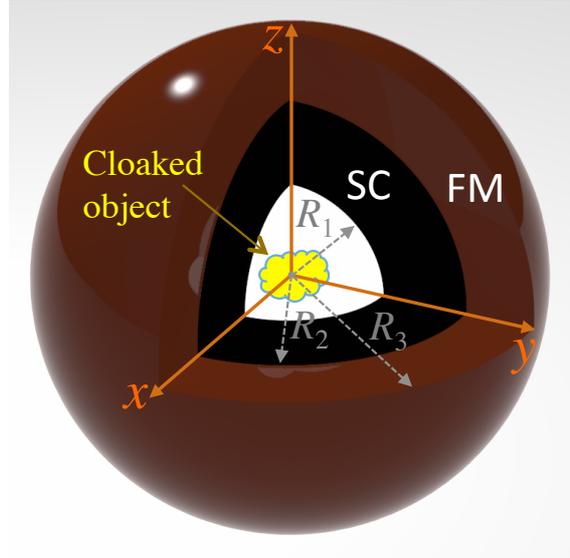

**Figure 1 | Schematic of the device.** The bilayer structure consists of an inner SC shell ($R_1 \leq r < R_2$), shown in black, and an outer FM shell ($R_2 \leq r \leq R_3$), shown in brown. The cloaked object (shown in yellow) is placed inside the cloaking region ($r < R_1$), shown in white. In our measurement, the *c*-axis of the YBCO's unit cell in the SC shell is defined along the z-axis, and the *ab*-lattice plane is parallel to the *xy* coordinate plane. In our experiment, we select $R_3 = 1.5R_2 = 15$ mm, which leads to $\mu_{FM} = 1.63$.

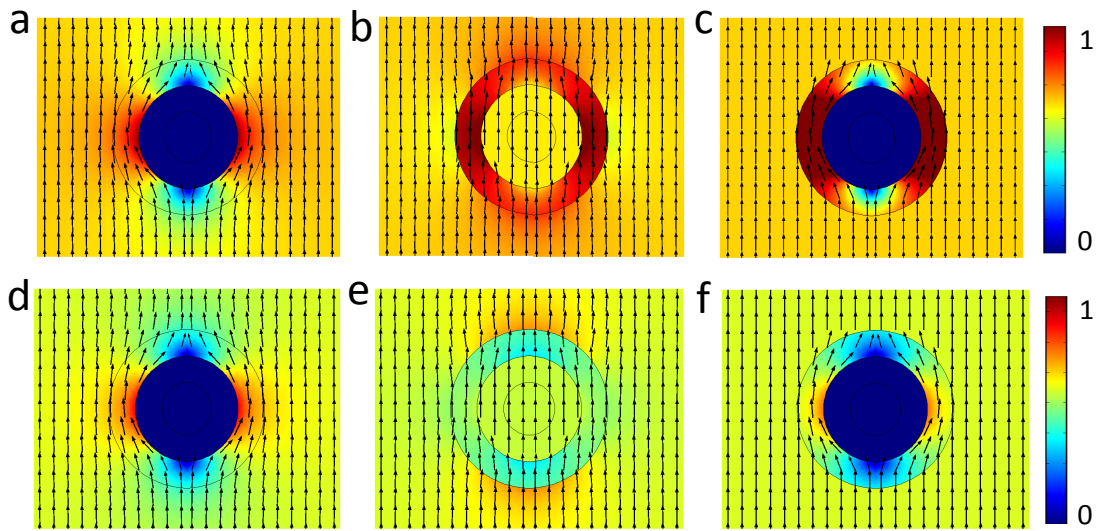

**Figure 2 | Simulation results. a-c,** the magnetic field patterns for the static field. **d-f,** snapshots of the magnetic field patterns for the time-harmonic field at 25 kHz. The samples are made of SC material only (**a** and **d**), FM material only (**b** and **e**) and the bilayer composite (**c** and **f**), respectively. The normalized magnetic field is represented by different colors and their directions are represented by the black arrow lines. For both static and dynamic cases, the bilayer sample shows no perturbation to the external magnetic field, and thus a perfect 3D cloaking is realized under the

quasi-static approximation.

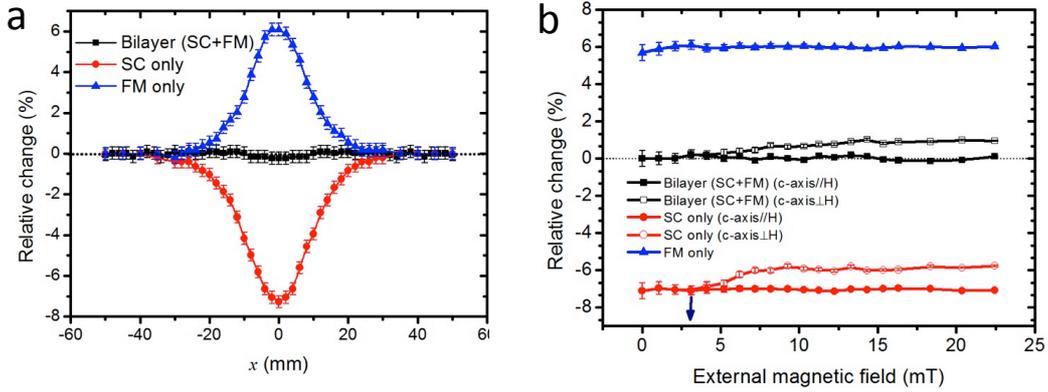

**Figure 3 | Measurement results for static fields. a,** relative change of the magnetic field measured for the bilayer sample (black squares) and two references with SC (red circles) and FM (blue triangles) shells only along the straight-line at $z = R_3 + 5$mm in the *xz* plane. In the measurement, a uniform external magnetic field of 2.5 mT is applied along the *z*-axis. **b,** the relative change of the magnetic field as a function of the strength of the applied magnetic field. The sample labeled "FM only" is measured along the *z*-axis and the samples labeled "SC only" are measured along both the *z*-axis (*c*-axis//H) and *x*-axis (*c*-axis⊥H). The anisotropic property of the lower critical field for YBCO single crystals leads to the "splitting" of the measured in-plane relative change at about 2.8 mT.

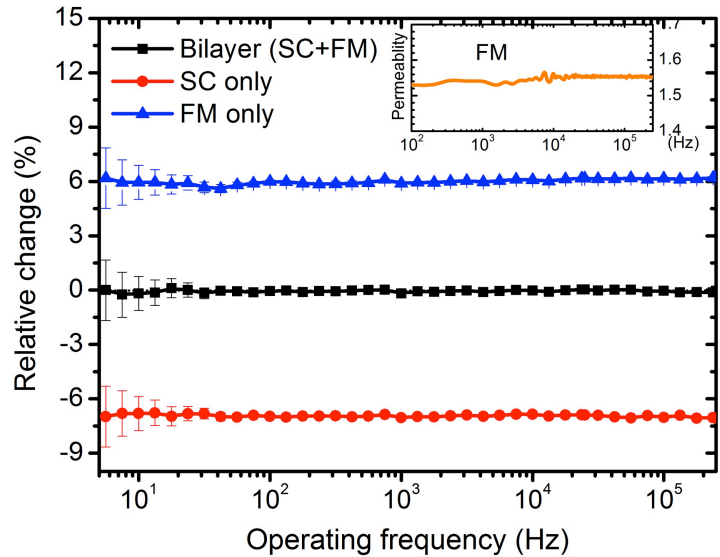

**Figure 4 | Measurement results for time-harmonic fields.** This displays the measured relative change of the magnetic field at a fixed point of $z = R_3 + 5$ mm and $x = y = 0$ when the operating frequency varies from 5 Hz to 250 kHz. Here, the magnetic field amplitude we apply is smaller than 1 mT. The intensity of the measured induction signal is frequency-dependent and has a large uncertainty error at lower frequencies, especially below 50 Hz. The samples of FM and SC shells only

have nearly constant values around 6% and 7%, respectively, while the bilayer structure has almost zero relative change in the measurement band. The inset plots the real permeability spectrum of the FM material measured at room temperature from 100 Hz to 250 kHz. The diluted composite has a negligible loss tangent (less than 0.01). The permeability spectrum at 77 K is estimated by multiplying by a factor proportional to the increase of the saturation magnetization due to the reduction in temperature.

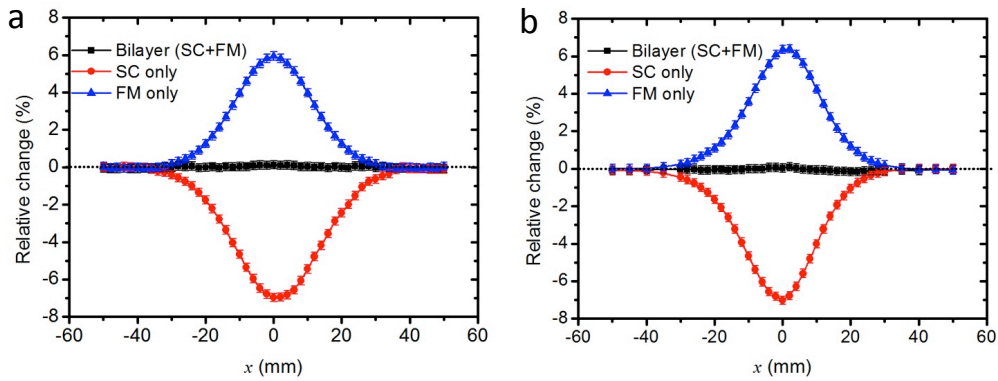

**Figure 5 | Measurement results at 25 kHz.** The measurement is conducted by scanning the prober over a fixed sample **a**) or moving the sample over a fixed probe **b**) along the straight-line at $z = R_3 + 5$mm in the $xz$ plane. In the second measurement, the magnetic field incident on the samples has a relative strength change of about 4% due to the spatial inhomogeneous distribution of the excitation field. The two different measurements yield similar results, indicating the robustness and isotropic field response of our samples.